\documentclass{article}

\usepackage[pdftex,colorlinks=true,plainpages=false,pdfpagelabels]{hyperref}

\def\beq{\begin{eqnarray}}
\def\eeq{\end{eqnarray}}
\def\bea{\begin{eqnarray*}}
\def\eea{\end{eqnarray*}}

\def\centeron#1#2{{\setbox0=\hbox{#1}\setbox1=\hbox{#2}\ifdim
\wd1>\wd0\kern.5\wd1\kern-.5\wd0\fi
\copy0\kern-.5\wd0\kern-.5\wd1\copy1\ifdim\wd0>\wd1
\kern.5\wd0\kern-.5\wd1\fi}}
\def\ltap{\;\centeron{\raise.35ex\hbox{$<$}}{\lower.65ex\hbox{$\sim$}}\;}
\def\gtap{\;\centeron{\raise.35ex\hbox{$>$}}{\lower.65ex\hbox{$\sim$}}\;}

\def\singleandthirdspaced{\baselineskip=\normalbaselineskip\multiply
    \baselineskip by 130\divide\baselineskip by 100}

\newcommand{\newc}{\newcommand}
\newc{\qbar}{{\overline q}}
\newc{\Kahler}{K\"ahler }
\newc{\deltaGS}{\delta_{\rm GS}}

\begin{document}
\begin{titlepage}
\begin{flushright}
{\large
SCIPP 10/07\\
}
\end{flushright}

\vskip 1.2cm

\begin{center}

{\LARGE\bf Axions in the Landscape and String Theory}

\vskip 1.4cm

{\large  Michael Dine$^{(a)}$, Guido Festuccia$^{(b)}$, John Kehayias$^{(a)}$ and Weitao Wu$^{(a)}$}
\\
\vskip 0.4cm
{(a) \it Santa Cruz Institute for Particle Physics and
\\ Department of Physics,
     Santa Cruz CA 95064  }\\
     {(b) \it School of Natural Sciences, Institute for Advanced Study\\
Einstein Drive, Princeton, NJ 08540}  \\

\vskip 4pt

\vskip 1.5cm

\begin{abstract}
While axions seem ubiquitous in critical string theories, whether they might survive in any string theoretic description
of nature is a difficult question.  With some mild assumptions, one can frame the issues in the case that there is an approximate
supersymmetry below the underlying string scale.  The problem of axions is then closely tied to the question of how moduli
are fixed.  We consider, from this viewpoint, the possibility that supersymmetry is broken at an intermediate scale, as in ``gravity
mediation," at a low scale, as in gauge mediation, and at a very high scale, to model the possibility that there is no low energy
supersymmetry.  Putative mechanisms for moduli fixing can then be systematically classified,
and at least for intermediate and high scale breaking, light axions appear plausible.  In the course of this work, we are lead to consider aspects of moduli fixing and supersymmetry breaking, and we revisit the possibility of very
large extra dimensions.

\end{abstract}

\end{center}

\vskip 1.0 cm

\end{titlepage}
\setcounter{footnote}{0} \setcounter{page}{2}
\setcounter{section}{0} \setcounter{subsection}{0}
\setcounter{subsubsection}{0}

\singleandthirdspaced

\section{Why String Theory Seems a Promising Setting to Solve the Strong CP Problem}
\label{axions}

>From the beginning, the Peccei-Quinn (PQ) solution to the strong CP problem had a troubling aspect:  why should the underlying theory obey a global symmetry
to an {\it extremely} good approximation.  Indeed, the approximation must be so good that any operator which would violate the symmetry
must be of very high dimension, even if its coefficient is scaled by inverse powers of the Planck scale.  As a result, to tackle the problem one requires a theory where
such ultraviolet questions can be meaningfully addressed.  String theory {\it does} appear to provide an answer:  there are often scalar fields, which, in perturbation
theory have no potential, and which couple to $F \widetilde F$ of candidate standard model gauge
fields \cite{wittenaxion}.  But, while suggestive, assessing the viability of the axion in string theory, and more
generally the axion itself (viewing string theory as a model for the axion phenomenon) requires understanding how supersymmetry is broken
and moduli are fixed.   If the non-perturbative effects which break supersymmetry also break the Peccei-Quinn symmetry, then the symmetry cannot
resolve
 the strong CP problem.  In the KKLT proposal \cite{kklt}, the most complete proposal for how moduli might be fixed in string theory, there is one seemingly natural axion candidate, the pseudoscalar field which is the partner of
the single K\"ahler modulus.  But this field gets a large mass, larger than the scale of supersymmetry breaking.  Indeed, the full supermultiplet gains a mass which
is approximately supersymmetric.  This has lead to pessimism about
obtaining QCD axions in a landscape \cite{bdg,donoghue}.  In this paper, we revisit this issue,
and discover that there are circumstances where axions of sufficient {\it quality} to solve the strong CP problem are plausible.

In order to assess the likelihood of light axions in string theory, with the present (limited) state of our understanding, it is necessary to
make some assumptions about the
underlying symmetries and dynamics.  Probably the most important of these is the scale of supersymmetry breaking.  We will focus here
on three possibilities:  intermediate scale breaking (ISB) of supersymmetry, as in ``gravity mediation" ($M_{int}\approx \sqrt{{\rm TeV}M_p}$), lower scale of supersymmetry breaking,
as in gauge mediation, and a scale much higher than the intermediate scale, as a model for a situation with no low energy supersymmetry.

In all three cases, if moduli are somehow fixed and are not extremely light, the partners of any very light axions must be fixed by {\it supersymmetry
breaking} dynamics.  So we wish to investigate whether moduli can indeed be fixed in such a fashion, while leaving over a Peccei-Quinn shift
symmetry of high quality.

The known axion candidates in string theory respect
certain shift symmetries.  Given our assumption that supersymmetry survives below
the fundamental scale, these lie in chiral multiplets, which we will refer to as ``axion multiplets:"
\beq
\tau_i \rightarrow \tau_i + 2\pi i n_i
\eeq
We choose here to take the axion to be the imaginary part of $\tau_i$,
\beq
\tau_i = s_i + i a_i.
\eeq
We will sometimes refer to $\tau_i$ and/or $s_i$ as ``axionic supermultiplets."

 If generated by short distance (e.g.~string scale
instanton) effects and if $s_i$ is large,
these necessarily appear in any superpotential
through terms of the form
\beq
W_\tau = e^{-m_i\tau_i},
\eeq
for some integers $m_i$.
We will assume that the axion decay constants are of order
the underlying fundamental scale, $M$,  which we will imagine ranges from $10^{15}$ GeV to the Planck scale (though at one point we will relax
this, when we consider the large extra dimension scenario of \cite{quevedo}).
We will assume that the continuous shift symmetry is broken by high scale ``instanton" effects of order $e^{-n\tau}$, with $n$ an integer.  The requirement
that the Peccei-Quinn symmetry is of high quality is then a constraint on the $s_i$'s.  To qualify as a QCD axion, for example, assuming that the fundamental scale is $10^{16}$ GeV,
 gives $s > 180$ or so.
 We will see, following ideas of \cite{Choi:2006za,acharya,raby} that there are quite plausible mechanisms which could account for the required $s_i$ for multiple axions.  The low energy dynamics of these moduli
 can be analyzed simply within the framework of low energy effective field theory.   Moreover, these additional moduli may themselves be the source of supersymmetry breaking and
 small, positive cosmological constant (c.c.).  The main difficulties with this framework are the familiar
ones of ISB:  flavor, the cosmological moduli problem
\cite{bkn,casasquevedo}, and the overshoot (or Brustein-Steinhardt \cite{bs})  problem.
The framework which we develop to study the dynamics of such
light axions allows a quite simple analysis of the  large extra dimension scenario of \cite{quevedo}.  We will see that the scenario is prone to a variety
 of instabilities, but, as has been discussed in the literature on this subject, it remains quite plausible that such stationary
 points exist.  Whether they are particularly common in a landscape is another matter, about which we can (and will) only speculate.

An alternative possibility is that the Peccei-Quinn symmetry arises as an accidental, approximate symmetry, linearly realized at scales well below $M$.
We will argue, following \cite{cdf}, that in the case of low energy supersymmetry breaking, cosmology probably necessitates such a picture.  The most likely way such a symmetry might arise
would seem to be the presence of discrete $R$ (and possibly other) symmetries.  This scenario does not suffer from flavor problems, or
 the usual cosmological moduli problem.  The question, in this context, is why the quality of the Peccei-Quinn symmetry is so high.

In this note, we will explore more generally the question of Peccei-Quinn symmetries in an underlying landscape.  In line with our
remarks above, we will consider various possible situations:
\begin{enumerate}
\item  Supersymmetry broken at an intermediate scale (ISB).
\item  Supersymmetry broken at a low scale, as in gauge mediation.
\item  Supersymmetry broken at a high scale, in part in order to model the absence of supersymmetry.
\end{enumerate}

We will see that the possibilities for moduli fixing, with large $s_i$, can be simply classified.  They include two mechanisms which fix a linear combination of
the $\tau_i$'s, while preserving an approximate supersymmetry:
\begin{enumerate}
\item  The small $W_0$ mechanism \cite{Choi:2006za,raby}.  Here $W_0$ is a constant term in the superpotential.  In this case, a linear combination of the $s_i$'s is fixed
in an approximately supersymmetric fashion, with the large value of $s_i$ being related to the small value of $W_0$.
At low energies. one linear combination of $\tau_i$'s can be integrated out, leaving a
theory with  a set of $\tau_i$'s, a constant superpotential, and with
some K\"ahler potential.  The (generalized) KKLT scenario fits within this framework.
\item  The ``Racetrack:''  Here, the hierarchy arises from a delicate balance between different exponentials in the superpotential.
Again, some linear combination of moduli are fixed, leaving other light fields which can play the role of the QCD axion (and
additional axions) \cite{acharya,raby}.  This theory, again, is described by a constant superpotential and some K\"ahler potential.
\end{enumerate}

Alternatively, {\it all} of the $\tau_i$'s can be fixed simultaneously with supersymmetry breaking.  This falls within the class of stabilization
mechanisms known as ``K\"ahler stabilization."

In the next section, we consider the small $W_0$ scenario.
We review why there is no light axion in the case of a single light modulus.  We explain that in the case of multiple light moduli, the effective low energy
 theory is very simple, and it can readily be seen that:
\begin{enumerate}
\item  Under plausible conditions, all of the $s_i$ moduli are stabilized by supersymmetry breaking.
\item  There can be multiple light axions, including the QCD axion, with the required high quality.
\item  There is a quite real possibility of breaking supersymmetry through the dynamics of the axionic moduli, without additional fields.
\end{enumerate}

In Section \ref{moregeneral}, we discuss racetrack and K\"ahler stabilization. Like the small $W_0$ case, in the case of the racetrack, there is typically one
modulus with mass large compared to $m_{3/2}$, while others have masses, again, of order $m_{3/2}$.  In the case of K\"ahler stabilization,
all of the moduli have masses of order $m_{3/2}$.

In Section \ref{susybreaking}, we discuss the question of whether supersymmetry can be broken
by the dynamics of the K\"ahler moduli themselves, without invoking antibranes or low energy dynamics on branes, or other phenomena.
We show that this is a logical possibility.  In Section \ref{largevolume}, we show that the Large Extra Dimension scenario of \cite{quevedo} fits
naturally in this framework.  The appearance of exponentially large volumes is readily understood, as well as potential instabilities, which we discuss.
In Section \ref{isb}, we briefly review two well-known issues associated with models of intermediate scale supersymmetry breaking:
flavor and cosmological moduli, and their relevance to the axion solution of the strong CP problem.

In Section \ref{gaugemediation}, we discuss axions in the frameworks of low energy gauge mediation and high scale supersymmetry
breaking.   Finally, in our conclusions, we provide a scorecard for different settings of the Peccei-Quinn solution of the strong
CP problem.

\section{Axions in the Small $W_0$ Scenario with Multiple K\"ahler Moduli}
\label{multiplekahler}

The KKLT proposal is a well-known realization of the small $W_0$ scenario.
Independent of the microscopic details, it can be summarized in a very simple
low energy Lagrangian.  At scales below the fundamental scale, there is a single chiral field, $\rho$, with superpotential
$$W = W_0 + \alpha e^{-\rho/b}$$
The K\"ahler potential is:
$$K = -3\ln(\rho + \rho^\dagger)$$
$W_0$ is a constant (as is $\alpha$).  The most important aspect of the KKLT analysis is the claim that $W_0$ can be
extremely small; this becomes the small parameter which permits self-consistent
approximations (it accounts, in particular, for large $\rho$).  With this assumption one can
solve sequentially for a stationary point of the potential.
First, study the equation for a supersymmetric minimum:
$${\partial W \over \partial \rho} + {\partial K \over \partial \rho} W = 0.$$
The minimum occurs for
\beq
\rho \sim - b \ln \vert W_0 \vert+ b\log(-\log|W_0|).
\label{rhoequals}
\eeq
At the minimum, the $W_0$ term in the superpotential dominates by a
power of $\rho$.

The low energy Lagrangian is a supergravity Lagrangian with superpotential
\beq
W \approx W_0.
\eeq
The mass of the components of $\rho$ is of order
\beq
m_\rho^2 \sim e^K \vert W_0\vert^2 \rho^2.
\eeq
In particular, if supersymmetry is broken in some fashion (e.g. anti-D branes, as suggested in KKLT,
or some low energy dynamics, perhaps associated with fields on branes), the mass of the modulus multiplet is larger than the gravitino mass by a factor of $\rho$; this justifies
solving the equation for $\rho$ first.

If this is all there is to the low energy theory, supersymmetry is unbroken and the cosmological
constant is negative.  KKLT, and subsequently others, have provided a variety of scenarios for how supersymmetry might be broken,
and for the origin of a substantial positive contribution to the vacuum energy.  The most popular of these involves anti-D3 branes\footnote{As often formulated,
this is puzzling, as it is not clear how supersymmetry is {\it spontaneously} broken; additional fields are necessarily required, e.g.~to provide the longitudinal component of the gravitino.  The resolution may lie in the fact that the effective
 theory is not actually four dimensional, due to an infinite tower of additional fields.},
 but one can alternatively imagine that there are simply some additional
interactions in the low energy theory which break supersymmetry \cite{dinekklt,otherskklt}.
Adopting such a model,
we can represent this through the presence of a field, $Z$, with a non-zero $F$ component.
Depending on the underlying details, this $F$ component, and the K\"ahler potential, may exhibit
non-trivial $\rho$ dependence.  For the moment, we will make no assumptions about the field $Z$; in Section \ref{susybreaking}, we will
explore the possibility that this field is itself one of the K\"ahler moduli.

In this theory, the fields in the $\rho$ multiplet all have mass of order
$\rho m_{3/2}$, and there is typically an additional multiplet whose components have masses of order $m_{3/2}$ (often including a pseudomodulus).
Whatever the detailed mechanism of supersymmetry breaking, the $\rho$ multiplet, including the pseudoscalar candidate axion, has an approximately
supersymmetric spectrum, and the axion is not suitable for solving the strong CP problem.

Thinking about the problem more generally,
it is clearly necessary that the partner of any would-be axion gain
mass only as a consequence of supersymmetry breaking \cite{cdf};
otherwise the axion is massive.
This requires that the superpotential, to a high degree of accuracy, be a function of only one linear combination of moduli,
\beq
W = W_0 + e^{-\tau/b}
\eeq
(we will assume $b > 1$, as in gaugino condensation); here $\tau = \sum n_i \tau_i$ (for IIB orientifolds of Calabi-Yau manifolds, the $\tau_i$'s would represent additional K\"ahler moduli).

At the same time, it is necessary that all of the moduli, $\tau_i$,
be fixed, in a fashion such that $e^{-\tau_i} \ll e^{-\tau/b}$,  More precisely, to obtain an axion suitable to solve the strong CP problem
(and more generally, to give rise to the axiverse of \cite{axiverse}), one requires that
\beq
W = W_0 + e^{-\tau/b} + A_i e^{-\tau_i} + \dots.
\label{tauw}
\eeq
$\tau$ is then fixed along the lines of KKLT, and the (real parts) of the other K\"ahler moduli are all fixed as a result of
the structure of the K\"ahler potential, in such a way that all additional terms in the superpotential
extremely small.

This may seem non-generic, and this was the point of view of \cite{cdf}.  But it is actually not difficult to see how this might occur\cite{Choi:2006za,raby}.
For the superpotential of eq.~(\ref{tauw}), one can solve the equation
\beq
D_\tau W = 0
\eeq
with
\beq
\tau \approx -b \log(\vert W_0 \vert).
\eeq
If the K\"ahler metric for $\tau$ vanishes at large $\tau$, as is typical of familiar supergravity and string constructions,
then $\tau$ is heavy, and can be integrated out.  The remaining light fields, $\tau_i$, are described by a theory with some
K\"ahler potential (a function of $(\tau_i + \tau_i^\dagger)$) and a constant superpotential.
Suppose that the K\"ahler potential has a stationary point at $\tau_i^0$,
\beq
\partial_i K = 0.
\eeq
With no other fields or dynamics, this point would correspond to a supersymmetric, AdS vacuum.  If there are additional dynamics
which break supersymmetry (as assumed, for example, in the KKLT model) giving small cosmological constant, then,
quite generally, the masses of the remaining $\tau_i$
fields receive a contribution to their masses-squared equal to $m_{3/2}^2$ from their mutual interactions.
Consider the potential:
\beq
V_{(\tau_i + \tau_i^\dagger)} = e^K \left(\vert W_0 \vert^2 K_i K^{i j} K_{j} - 3 \vert W_0 \vert^2 \right),
\eeq
where we have noted that it is only the combination $(\tau_i +
\tau_i^\dagger)$ that appears in $K$ and thus the potential.  Differentiating
twice and using the fact that the $K_i$'s vanish at the stationary
point, indeed yields $-K_{ij}|W_0|^2e^{K(\tau_i^0 + \tau_i^{0\dagger})}$.

Whether this is the entire contribution to the masses of these moduli depends on the nature of supersymmetry breaking.  Suppose, for example,
that supersymmetry is broken by the $F$ component of a chiral field, $X$, through an additional term in the superpotential $fX$.
Then, depending on the structure of the K\"ahler potential for $X$, there can be additional contributions to the masses of the $\tau_i$ fields.
Terms such as $X^\dagger X/(\tau_i + \tau_i^\dagger)$ respect the shift symmetry, and contribute to the masses, potentially with either sign,
an amount of order $m_{3/2}$.  So whether these points are actually local minima of the potential depends on such details.

The procedure of integrating out the heavy modulus can be illustrated by a simple example.
We will use the language of the IIB theory.
First, suppose we consider a version of KKLT with two K\"ahler moduli, $\rho_1$ and $\rho_2$, and with superpotential
\beq
W = W_0 - A e^{-{ \rho_1 +  \rho_2 \over b}} + fX
\eeq
and K\"ahler potential:
\beq
K = -\frac{3}{2} \log \left ( (\rho_1 + \rho_1^\dagger) (\rho_2 + \rho_2^\dagger) \right ) + K(X,X^\dagger).
\eeq
Here, for simplicity, $K(X,X^\dagger)$ is such that it gives rise to a minimum for $X$ near the origin, e.g.
\beq
K = X^\dagger X + {1 \over m^2} X^\dagger X X^\dagger X.
\eeq
Such a superpotential could arise in the presence of gaugino condensation in a sector with gauge coupling
\beq
\rho_0 =\rho_1 + \rho_2
\eeq
and $\beta$-function
$b/3$.
As in the original KKLT model, this model has an approximate, supersymmetric stationary point at
\beq
\rho_1 = \rho_2 = x \approx { b \over 2} \log(\vert W_0 \vert).
\eeq
 $\rho_0$ has mass-squared of order $\rho_0^2\vert W_0\vert^2$, and can be integrated out, leaving an effective theory for
 \beq
 \Psi = \rho_1-\rho_2
 \eeq
 with K\"ahler potential and superpotential
 \beq
 K = {1 \over \rho_0^2} (\Psi + \Psi^\dagger)^2 + \dots + K(X,X^\dagger)~~~~W = W_0 + fX.
 \eeq
 For suitable adjustment of $W_0$ and $f$, as suggested in \cite{kklt,bp}, the cosmological
constant can be arbitrarily small.  In this case,
 the minimum of the potential for ${\rm Re}~\Psi$ lies at the origin, and the $\Psi$ field
has mass of order $m_{3/2}$.
It is easy to consider more realistic K\"ahler potentials and follow through the same procedure.

How light is the axion?  High energy non-perturbative effects are of order $e^{-\rho_0}$.  So if $b$ is of order, say $4$, and
\beq
e^{-\rho_0/b} M_p = 100 ~{\rm TeV}
\eeq
then
\beq
e^{-\rho_0} = 10^{-52}.
\eeq
This gives an axion of high enough ``quality" \cite{cdf} to account for the axion of QCD.  Of course, factors of two in the exponent can make
an appreciable difference in one direction or the other.

It is natural, in this low energy picture, to ask whether the field $X$ is necessary; could the low energy theory for some set of fields, $\Psi_i$,
with constant superpotential, be responsible for supersymmetry breaking?  We will explore this question in Section \ref{susybreaking}.

\section{Other Stabilization Mechanisms:  Racetrack Models and K\"ahler Stabilization}
\label{moregeneral}

The essential ingredients in our analysis above were:
\begin{enumerate}
\item  The presence, at some low energy scale, of multiple axion superfields, $\tau_i = s_i + i a_i$, invariant under a discrete shift symmetry.
\item  As a consequence of (1), for large $s_i$, holomorphic quantities depend on $\tau_i$ as $e^{-n_i\tau_i}$.  It is necessary
that $\rho_i$ be large if the corresponding axion is to be light.
\end{enumerate}
The existence of this small parameter for the would-be axion is particularly important.
 In the KKLT scenario, the large size of $\rho$  is correlated with the small parameter $W_0$; the large value of some linear
 combination of $\rho_i$'s is determined
 by supersymmetric dynamics; supersymmetry breaking determines the relative values of the different
 moduli.  We will take this as our definition of the KKLT scenario (in this sense, it applies to the proposals in \cite{acharya,raby}).

 There are other ways we might imagine a small
parameter could
come about, which would allow for the possibility of light axions.  First, there is the possibility that one obtains a large value
of all of the $\rho_i$'s through dynamics which break supersymmetry.
This would be a realization of ``K\"ahler stabilization'' \cite{kahlerstabilization}.
In this context, there would again be a superpotential which is a function of a linear combination of the moduli.  The K\"ahler potential of the theory
would give rise to supersymmetry breaking and a large expectation value for this modulus; each of the other moduli would similarly be large.  The spirit of the K\"ahler stabilization
hypothesis is that, despite the fact that the theory is not weakly coupled, certain holomorphic quantities, the factors $e^{-\rho_i}$ in this case, are extremely small.  Whether
this actually happens in string theory is an open question.

Another possibility is that the low energy theory for the $\tau_i$'s (the theory obtained after integrating out the complex structure moduli
in the case of KKLT) contains no constant
in the superpotential, perhaps due to an unbroken, discrete $R$ symmetry.  The superpotential for $X$ is would be generated by non-perturbative
effects, e.g.~as in retrofitted models \cite{retrofitted,dk}.  These effects would break the $R$ symmetry, generating the requisite expectation value of the superpotential.
Fixing of the K\"ahler moduli, with remaining axions, follows as in the previous section.

So arguably the appearance of axions in such a picture is robust.  So far, we have assumed that $m_{3/2} \sim ~{\rm TeV}$.  If we are not wedded to supersymmetry
as a solution of the hierarchy problem, we can consider the case of much larger $m_{3/2}$; then we have a model for axions
 without low energy supersymmetry.  What is mainly required is that $e^{-\rho_0/b}$ still be hierarchically small,
 so that plausible powers of this parameter can account for the quality of the QCD axion.
 Indeed, well known cosmological considerations \cite{bkn,casasquevedo} require
that the scale be $50$ TeV or larger; we will review these in Section \ref{isb}.

Lower $m_{3/2}$ would arise, for example, in gauge mediation.  Here it is problematic that the partner of the axion must be fixed by
supersymmetry breaking dynamics.  If it's interactions are Planck suppressed, this particle leads to an untenable cosmology.  To obtain
axions, one either requires that the scale of interactions between the axion multiplet and the hidden sector is significantly lower than
$M_p$, or that there are no light moduli, and the Peccei-Quinn symmetry is linearly realized for some range of scales
in the low energy theory.
We will discuss these issues in Section \ref{gaugemediation}.

\section{Supersymmetry Breaking With Multiple Moduli}
\label{susybreaking}

We have seen in the previous section that if there are additional
light fields which break supersymmetry, we can readily understand how all of the K\"ahler moduli are fixed in a way that non-perturbative
corrections to the axion potentials are small.
In this section, we discuss the possibility that the additional K\"ahler moduli are themselves responsible for supersymmetry breaking.  In general,
we are dealing with a theory with a constant superpotential and some K\"ahler potential.   It is easy to see that it is possible, in principle, for supersymmetry to be broken
in this situation.

Near any would-be stationary point of the potential, we still expect to find one heavy field, $\Phi$, and several light fields, $\rho_i$.  These fields
are distinguished by the invariance of the theory under discrete shift symmetries.    Integrating out
$\Phi$, the low energy theory will again be described by a constant superpotential, $W_0$, and a K\"ahler potential for $\rho_i$.  We still have good,
approximate shift symmetries,
\beq
\rho_i \rightarrow \rho_i + 2 \pi i n_i.
\eeq
We can ask whether this theory can break supersymmetry.  In fact, under rather general conditions, this low energy theory does exhibit supersymmetry
{\it preserving} and supersymmetry {\it breaking} stationary points.
The supersymmetry preserving points are AdS, and satisfy the
Breitenlohner-Freedman bound \cite{Breitenlohner1982,Breitenlohner1982a}.
The supersymmetry breaking points have lower (i.e.~more negative) cosmological constant.  So if these states are obtained in any sort of systematic
approximation, additional dynamics are required to account for a small, positive cosmological constant.

If the function $K(\rho_i + \rho_i^\dagger)$ has a stationary point:
\beq
\partial_i K = 0 ~\forall i; ~\rho_i = \rho_i^0
\eeq
then supersymmetry is unbroken at this point, with negative potential,
\beq
V_0 = - 3e^{K(\rho^0)}\vert W_0 \vert^2 ~~~
\eeq
while each of the $(\rho_i + \rho_i^\dagger)$'s are tachyonic, with
\beq
m^2 = - 2e^{K(\rho^0)}\vert W_0 \vert^2.
\eeq
The Breitenlohner-Freedman bound is satisfied
\beq
-\frac{9}{4} \le m^2R^2 = -2
\eeq
so these configurations describe stable AdS vacua.  However, because the curvature of the potential is negative at this point, and because for large $\rho_i$ the potential typically tends to zero, we expect that the potential exhibits AdS supersymmetry breaking solutions.

\subsection{Simple Models which Break Supersymmetry}

This is illustrated by a simple example (we will also discuss the large volume solutions of \cite{quevedo} in Section \ref{largevolume}).
Rather than consider a K\"ahler potential with logarithmic behavior at $\infty$, take
$$K = \frac{1}{2}(\rho + \rho^\dagger)^2.$$
This admits a supersymmetric solution at $(\rho + \rho^\dagger) = 0$, and a
non-supersymmetric solution at
\beq
(\rho + \rho^\dagger) = \pm 1.
\eeq
The non-supersymmetric solution, indeed, has lower energy.

Typically, when reliable computations are possible, this is problematic; the supersymmetry-breaking solutions are AdS.  So additional dynamics are required to account
for the observed small, positive cosmological constant.

But it is not at all clear that the underlying microscopic theory should be weakly coupled, and even if it is, its K\"ahler potential might be more complicated
than we have contemplated above.  We can ask whether it is logically possible, in a theory with constant $W$, to have broken supersymmetry and
vanishing cosmological constant.  In fact it is, as can be seen by considering a theory with a single scalar field, $\psi$, invariant under shifts.

We will suppose that the would-be minimum lies at $\psi=0$, and expand the K\"ahler potential about this point:
\beq
K(\Psi + \Psi^\dagger) \equiv K(\psi) = k_0 + k_1 \psi + {k_2 \over 2} \psi^2 + {k_3 \over 3} \psi^3 + {k_4 \over 4} \psi^4.
\eeq

We then require that the constants $k_i$ satisfy the following conditions:
\begin{enumerate}
 \item
 The potential has a stationary point at $\psi =0$.
 \item
 The potential is a {\it minimum} at $\psi = 0$.
 \item
 The kinetic terms for $\psi$ have a sensible sign at $\psi =0$:  $k_2 > 0$.
 \item  Supersymmetry is broken at $\psi =0$:  $k_1 \ne 0$.
 \item  The cosmological constant is (nearly) zero at $\psi =0$:  $k_1^2 = 3k_2$.
 \label{ccpoint}
 \end{enumerate}
 It is easy to see that these conditions can be simultaneously satisfied.

\section{The Exponentially Large Volume Scenario}
\label{largevolume}

The authors of  \cite{quevedo} have put forth a scenario for the flux landscape in which the volume is ``exponentially large.''  We will see that their scenario
can be analyzed within the framework we have put forward here.  There is a parameter, $W_0$ (which should be small, but need not be extremely
small).  Integrating out one modulus supersymmetrically leads to a low energy theory for the remaining
moduli with constant superpotential and a particular K\"ahler potential.  The leading terms in the potential, for this particular
K\"ahler potential, cancel, leaving terms which vary logarithmically with the light modulus (this logarithmic variation
is just the logarithmic modulus dependence familiar in KKLT).  These terms are of the same order, for large
values of the modulus, as the first subleading $\alpha^\prime$ corrections.  The competition of these terms then
allows for exponentially large solutions.  The controlling, small parameter for their analysis, we will
see, is $g_s$, the string coupling.  In this setup, it is easy to understand why a modulus exponential in $1/g_s$ naturally arises in the leading
order analysis.  It is also clear  that this solution is potentially unstable.  It is crucial to understand the form of stringy perturbative
corrections to the K\"ahler metric for various fields.  Some work has
been done on this question \cite{haack,quevedoconloncorrections},
and further investigations will be reported elsewhere \cite{kehayiaswutoappear}.

Without reviewing all of the details of the large dimension models, suffice it to say that, like the models considered here, there are multiple
K\"ahler moduli (in the examples they analyze in detail, 2).  Following their notation, we will refer to these
as $\tau_4,\tau_5$.  The K\"ahler potential has a different form than that we have studied up to now:
\beq
K = -2 \ln \left ({\cal V} + {\xi \over 2g_s} \right ).
\eeq
where
\beq
{\cal V} = \tau_5^{3/2} - \tau_4^{3/2} \approx \tau_5^{3/2} \left(1 - \left ({\tau_4 \over \tau_5} \right )^{3/2} \right),
\eeq
and where $\xi$ is a numerical constant.
The superpotential can be taken to be (our notation is not identical to that of \cite{quevedo}, and our form differs slightly, by redefinitions of fields).
\beq
W = W_0  + A e^{-\rho_4}.
\eeq
Here $\rho_4 = \tau_4 + i a$, i.e.~$\rho_4$ is the superfield whose lowest component is $\tau_4$ plus an additional axion field.

Now, as in our previous analyses, $\tau_4$ is the heavy field, and we should integrate it out, solving its equations of motion.  The equation $D_{\tau_4} W=0$
gives:
\beq
\tau_4[\tau_5] \approx -  \ln\left(W_0/\tau_5^{3/2}\right).
\eeq
$\tau_5$ is the light field.  To leading order in $1/\tau_5$, the K\"ahler potential for $\tau_5$ is:
\beq
K = -3\ln(\tau_5) + 2 \frac{\tau_4[\tau_5]^{3/2}}{\tau_5^{3/2}} - {\xi \over g_s \tau_5^{3/2}}.
\label{approximatek}
\eeq
where we have explicitly indicated that $\tau_4$ should be thought of, here, as a function of $\tau_5$.
The superpotential for $\tau_5$ is approximately $W_0$, as in our earlier examples.

Because of the so-called ``no-scale structure" of the K\"ahler potential, the leading terms in the potential, of order $m_{3/2}^2 M_p^2 = {1 \over \tau_5^3}{\vert W_0 \vert^2 \over M_p^2}$
cancel.
The next order terms, generated by the second and third terms in eq.~(\ref{approximatek})
each behave as $\tau_5^{-9/2}$, {\it up to logarithms}.  More precisely the potential behaves as a function of powers  of $\log(\tau_5)$
times $\tau_5^{-9/2}$.  It is this feature which leads to a stationary point at exponentially large $\tau_5$.

In the approximation that $\tau_4$
is large, the calculation of the potential greatly simplifies.  In particular, in taking derivatives of $K$, terms obtained
by differentiating $\tau_4$ are suppressed.
\beq
K^\prime \approx - {3 \over \tau_5} \left ( 1 + {\tau_4^{3/2}
    \over \tau_5^{3/2}} - {1 \over 2} {\xi \over g_s \tau_5^{3/2}} \right )
\eeq
\beq
(K^{\prime \prime})^{-1} \approx {\tau_5^2 \over 3} \left ( 1 - {5 \over 2} {\tau_4^{3/2} \over \tau_5^{3/2}} + {5 \over 4} {\xi \over g_s \tau_5^{3/2}} \right ).
\eeq
Correspondingly, the potential is approximately: \beq
V = -3 |W|^2\left ({1 \over 2} {\tau_4^{3/2} \over \tau_5^{3/2}} - {1 \over 4} {\xi \over g_s \tau_5^{3/2}} \right ) {1 \over \tau_5^3}
\eeq
Since both terms exhibit the same $\tau_5^{-9/2}$ power law dependence, the competition of the logarithmic dependence of the first term and the (large, for small $g_s$) constant in the second term gives rise to an exponentially large solution:
\beq
\tau_5 \approx e^{(\xi/g_s)^{2/3}}
\eeq
This is the result found in \cite{quevedo}.

There are now several issues.  First, the result is crucially dependent on the structure of the leading terms in the K\"ahler potential.  Corrections in $g_s$ might spoil
this.  For a smooth manifold, one might expect these corrections to be suppressed by powers of the compactification radius.  This follows from the fact that at distances
small compared to $R$, the theory is ten dimensional, so contributions to loops from high Kaluza-Klein modes would be essentially ten dimensional, in which
case supersymmetry forbids corrections to kinetic terms.  Still, the power must be rather large, and one might worry, in addition, that an orientifold
is not a smooth manifold.  These questions have been studied in \cite{haack} and \cite{quevedoconloncorrections}, who argue that, while such corrections are present,
they are such that the first subleading corrections to the {\it potential} cancel as well, and the result is stable.  This question will be studied further
in \cite{kehayiaswutoappear}.

A second issue is that the cosmological constant is negative in these states\footnote{Note that if the cosmological constant were positive, this would be particularly
  problematic, since by assumption corrections are small.   Additional supersymmetry breaking dynamics might be expected to add only
  additional positive contributions, so some more
  significant modification of this structure would be required.}.  By assumption, the calculations are reliable in this limit,
so additional dynamics are required in order to understand a small, positive vacuum energy.  These have the potential to further destabilize
the vacuum.  As a model, add a chiral field $X$ to the theory, with superpotential
\beq
\delta W = fX
\eeq
and K\"ahler potential adjusted so that the $X$ potential has a stationary point at the origin.  We require that $f$ be such that it cancel the vacuum
energy $-V_0$.  This requires
\beq
\vert f^2 \vert \sim {\vert W_0 \vert^2 \log(\tau_5)^{1\over 2}\over (\tau_5^0)^{3/2}}
\eeq
where $\tau_5^0$ is the value of $\tau_5$ near the minimum.

For such a value of $f$, this is a small perturbation on the $\tau_5$ potential and the value of
$\tau_5$  at the minimum $\tau_5^0$ changes by an order one multiplicative constant with respect to the $f=0$ case.  The viability of the large dimension solution critically depends on this tuning of $f^2$.  If $f^2$ were too large, say by a factor of $100$, the ``corrections" to the
$\tau_5$ potential would overwhelm those considered in \cite{quevedo}.  One would be driven back to the small $W_0$ (KKLT)
scenario.  In a landscape, then the question might be something like:  how common are exponentially large dimensions
(requiring tuning of $f^2$) vs.~very small $W_0$?

Even if we suppose that the cosmological constant can be explained while still obtaining an exponentially large volume solution,
there are other issues.
Examining the potential and the kinetic terms, we see that the mass of $\tau_5$ is small compared to $m_{3/2}$;
\beq
m_{\tau_5}^2 \approx m_{3/2}^2 \tau_5^{-3/2}
\eeq
due, again, to the ``no scale" cancellations.  As has been widely discussed in the literature, this is potentially problematic cosmologically.
To deal with this issue, it has been suggested \cite{conlon} that the scale, $m_{3/2}$ is large, and that matter fields are light (by powers of $\tau_5$ compared
to $m_{3/2}$)
due to a no scale structure.

It is important that this feature survive at the quantum level.
Examination of low energy diagrams suggests that this hierarchy can be
stable \cite{conlonstability},
but this is worthy of further investigation.

Overall, then, the low energy effective field theory approach leads to a clear understanding of the exponentially large volume scenario.  We see that the
stability of the large volume solution is critically dependent on the precise structure of the K\"ahler potential and the dynamics
responsible for cancellation of the cosmological constant.  Arguably, in a broad range of circumstances,
corrections to the K\"ahler potential may well be small enough to support a large volume solution.  But in the regime where the analysis is potentially
reliable, it predicts a negative cosmological constant, so additional dynamics are required.  The dynamics responsible for tuning the cosmological constant
seems to sustain the large extra dimension solution precisely (and only) as a consequence of this tuning.

\section{General Issues in ISB}
\label{isb}

Models with ISB raise at least two well-known issues.  The first is the flavor problem.  We have nothing new to add on this subject except to note that the seeming
ease with which one might generate a suitable PQ symmetry is perhaps good reason to reconsider this issue, and the various solutions (flavor symmetries,
features of particular regions of the moduli space) which have been proposed.

The second issue is the moduli problem.  As stressed in \cite{bdgraesser}, in supersymmetry axions are inevitably associated with a cosmological
moduli problem, which is parametrically more severe than the usual axion energy density problem.  There are basically two proposed solutions.
First, moduli might sit at enhanced symmetry points.  But this is incompatible with the moduli partners as axions.  Second, the moduli might be sufficiently
massive that the temperature after their decays is high enough to
restart nucleosynthesis \cite{bkn,casasquevedo}.   This corresponds to moduli masses
 (and presumably the gravitino mass) larger than $10$ TeV.  As usual, this poses a fine tuning problem (conceivably
one whose solution could be understood anthropically).
One also must account for limited production of stable particles (LSP's) in moduli decays.
But it has the effect that it significantly relaxes the constraint on the axion decay constant, readily
allowing axion decay constants of $10^{14}$ GeV or perhaps somewhat larger, without invoking a small misalignment angle.

Related to these questions is the question of overshoot \cite{bs}.  Models of the type discussed here will suffer from this
difficulty; perhaps the most plausible solution is a modest tuning of initial conditions, as discussed in \cite{susskindovershoot}.  This tuning might
plausibly have an anthropic origin.
One can ask whether the problem takes a different form in the presence of multiple moduli.  There are now several fields each of whose initial
conditions must, to a similar degree, be tuned.  The severity of this
tuning might disfavor the ``axiverse" scenario \cite{axiverse}.  On the other hand,
in a landscape, it is conceivable that there are many more ``states" with larger numbers of K\"ahler moduli,
and this effect might overwhelm the effects of tuning, favoring an axiverse.   These issues will be discussed in a subsequent
publication \cite{wutoappear}.

\section{Axions in Low Scale (Gauge Mediation) and High Scale Supersymmetry Breaking}
\label{gaugemediation}

By low scale supersymmetry breaking, we mean $m_{3/2} \ll$ TeV.  Generally,
for such low scales, assuming Planck scale couplings for the moduli as would naively
arise in the KKLT scenario, for example, these fields are cosmologically problematic.
They quickly come to dominate the energy of the universe, and at the same time they
are quite light, decaying long after nucleosynthesis.

So if supersymmetry is broken at low scales,
then we have to assume that, in fact, there are no light moduli at scales
just below the fundamental scale.  In that case, any would-be axions must arise from a Peccei-Quinn symmetry which is linearly
realized below the fundamental scale.  This possibility is explored in \cite{cdf}.  It is not difficult to write such models, accounting for
the quality of the Peccei Quinn symmetry through discrete symmetries.  These symmetries, however, are
rather intricate and it is not clear why such a structure would be generic.    In this framework, it is natural for $f_a$ to be
$10^{11}-10^{12}$ GeV,
as this scale is readily connected to some messenger scale, and also because it tends to ameliorate the quality problem.

This framework avoids the ISB moduli problem, and, assuming a gauge mediated structure, also avoids problems of flavor changing
currents.  But the existence of a high quality PQ symmetry is a puzzle.


Given our limited knowledge of string theories without supersymmetry, we can consider, instead, possible vacua in which the supersymmetry
breaking is well below the fundamental scale, but not so low as to resolve the hierarchy problem.    We can again take KKLT as a model
for moduli stabilization.  There seems no difficulty in accounting for an axion in this framework.
Needless to say, the flavor problems of ISB are alleviated in such a regime, as are the cosmological issues connected with
moduli.  So axions would seem particularly plausible in the absence of low energy supersymmetry.
But as the supersymmetry breaking scale approaches the fundamental scale, the problem of the axion quality becomes progressively
more severe, suggesting that some degree of low energy supersymmetry might have something to do with the solution
of the strong CP problem.

\section{Outlook and Scorecard}

String theory provides the most promising setting for the PQ solution of the strong CP problem.
We have seen that:
\begin{enumerate}
\item  In theories with approximate supersymmetry below the fundamental scale, the superpartners of would-be axions must be fixed by supersymmetry breaking
dynamics.
\item  In the small $W_0$ (or racetrack or K\"ahler stabilization) scenario(s), with multiple K\"ahler moduli, one can readily understand the presence of axions of high quality, from quite generic low energy
effective actions.
\item  In the small $W_0$ scenario, these additional moduli can readily, themselves, be responsible for supersymmetry breaking.
\item  ISB suffers from the standard problems of flavor and cosmological moduli.  The Brustein-Steinhardt problem may be more severe in such a picture.
\item  In low energy supersymmetry breaking, any PQ symmetry (responsible for the QCD axion and possibly
 other light pseudoscalars) should be linearly realized below the fundamental scale; the breaking should be visible within
the low energy theory.  Flavor and moduli are not severe problems, but understanding the quality of the PQ symmetry is more challenging than in
IMB.
\item  As a model for the absence of low energy supersymmetry, we can simply take the supersymmetry breaking scale large (tuning the weak scale); suitable axions
remain highly plausible, and many of the problems of ISB are ameliorated.
\end{enumerate}

>From these observations, it seems that axions are a quite plausible outcome of a landscape picture for understanding the laws of nature.  ISB seems the most plausible
setting.  If the Peccei-Quinn symmetry is non-linearly realized below the fundamental scale, it is hard to see how to adequately protect the axion without low energy (but not
necessarily weak scale) supersymmetry.  If it is linearly realized, either in models of low energy supersymmetry breaking or in models without low energy supersymmetry,
intricate discrete (gauged) symmetries seem required to account for
the high degree of axion quality.
\vspace{2pc}

\noindent
{\bf  Acknowledgements}
Conversations with Tom Banks and Leonard Susskind are gratefully
acknowledged.  This work was supported in part by the U.S.~Department
of Energy.  M.~Dine thanks Stanford University and the Stanford
Institute for Theoretical Physics for a visiting faculty appointment
while much of this work was performed.  J.~Kehayias thanks Mark
Goodsell for useful conversations and references while at Susy10.
G.~Festuccia is supported by NSF grant NSF PHY-0969448. Any opinions,
findings, and conclusions or recommendations expressed in this
material are those of the authors and do not necessarily reflect the
views of the National Science Foundation.

\bibliographystyle{utphys}
\bibliography{axions_landscape_refs}

\providecommand{\href}[2]{#2}\begingroup\raggedright\begin{thebibliography}{10}

\bibitem{wittenaxion}
E.~Witten, ``{Some Properties of O(32) Superstrings},''
  \href{http://dx.doi.org/10.1016/0370-2693(84)90422-2}{{\em Phys.Lett.}
  {\bfseries B149} (1984) 351--356}.

\bibitem{kklt}
S.~Kachru, R.~Kallosh, A.~D. Linde, and S.~P. Trivedi, ``{De Sitter vacua in
  string theory},'' \href{http://dx.doi.org/10.1103/PhysRevD.68.046005}{{\em
  Phys. Rev.} {\bfseries D68} (2003) 046005},
\href{http://arxiv.org/abs/hep-th/0301240}{{\ttfamily arXiv:hep-th/0301240}}.

\bibitem{bdg}
T.~Banks, M.~Dine, and E.~Gorbatov, ``{Is there a string theory landscape?},''
  \href{http://dx.doi.org/10.1088/1126-6708/2004/08/058}{{\em JHEP} {\bfseries
  0408} (2004) 058}, \href{http://arxiv.org/abs/hep-th/0309170}{{\ttfamily
  arXiv:hep-th/0309170 [hep-th]}}.

\bibitem{donoghue}
J.~F. Donoghue, ``{Dynamics of M theory vacua},''
  \href{http://dx.doi.org/10.1103/PhysRevD.69.106012,
  10.1103/PhysRevD.69.129901}{{\em Phys.Rev.} {\bfseries D69} (2004) 106012},
  \href{http://arxiv.org/abs/hep-th/0310203}{{\ttfamily arXiv:hep-th/0310203
  [hep-th]}}.

\bibitem{quevedo}
J.~P. Conlon, F.~Quevedo, and K.~Suruliz, ``{Large-volume flux
  compactifications: Moduli spectrum and D3/D7 soft supersymmetry breaking},''
  {\em JHEP} {\bfseries 08} (2005) 007,
\href{http://arxiv.org/abs/hep-th/0505076}{{\ttfamily arXiv:hep-th/0505076}}.

\bibitem{Choi:2006za}
K.~Choi and K.~S. Jeong, ``{String theoretic QCD axion with stabilized saxion
  and the pattern of supersymmetry breaking},''
  \href{http://dx.doi.org/10.1088/1126-6708/2007/01/103}{{\em JHEP} {\bfseries
  01} (2007) 103},
\href{http://arxiv.org/abs/hep-th/0611279}{{\ttfamily arXiv:hep-th/0611279}}.

\bibitem{acharya}
B.~S. Acharya, K.~Bobkov, and P.~Kumar, ``{An M Theory Solution to the Strong
  CP Problem and Constraints on the Axiverse},''
\href{http://arxiv.org/abs/1004.5138}{{\ttfamily arXiv:1004.5138 [hep-th]}}.

\bibitem{raby}
K.~Bobkov, V.~Braun, P.~Kumar, and S.~Raby, ``{Stabilizing All Kahler Moduli in
  Type IIB Orientifolds},''
\href{http://arxiv.org/abs/1003.1982}{{\ttfamily arXiv:1003.1982 [hep-th]}}.

\bibitem{bkn}
T.~Banks, D.~B. Kaplan, and A.~E. Nelson, ``{Cosmological Implications of
  Dynamical Supersymmetry Breaking},''
  \href{http://dx.doi.org/10.1103/PhysRevD.49.779}{{\em Phys. Rev.} {\bfseries
  D49} (1994) 779--787},
\href{http://arxiv.org/abs/hep-ph/9308292}{{\ttfamily arXiv:hep-ph/9308292}}.

\bibitem{casasquevedo}
B.~de~Carlos, J.~A. Casas, F.~Quevedo, and E.~Roulet, ``{Model independent
  properties and cosmological implications of the dilaton and moduli sectors of
  4-d strings},'' \href{http://dx.doi.org/10.1016/0370-2693(93)91538-X}{{\em
  Phys. Lett.} {\bfseries B318} (1993) 447--456},
\href{http://arxiv.org/abs/hep-ph/9308325}{{\ttfamily arXiv:hep-ph/9308325}}.

\bibitem{bs}
R.~Brustein and P.~J. Steinhardt, ``{Challenges for superstring cosmology},''
  \href{http://dx.doi.org/10.1016/0370-2693(93)90384-T}{{\em Phys. Lett.}
  {\bfseries B302} (1993) 196--201},
\href{http://arxiv.org/abs/hep-th/9212049}{{\ttfamily arXiv:hep-th/9212049}}.

\bibitem{cdf}
L.~M. Carpenter, M.~Dine, G.~Festuccia, and L.~Ubaldi, ``{Axions in Gauge
  Mediation},'' \href{http://dx.doi.org/10.1103/PhysRevD.80.125023}{{\em
  Phys.Rev.} {\bfseries D80} (2009) 125023},
  \href{http://arxiv.org/abs/arXiv:0906.5015}{{\ttfamily arXiv:arXiv:0906.5015
  [hep-th]}}.

\bibitem{dinekklt}
M.~Dine, ``{The intermediate scale branch of the landscape},'' {\em JHEP}
  {\bfseries 01} (2006) 162,
\href{http://arxiv.org/abs/hep-th/0505202}{{\ttfamily arXiv:hep-th/0505202}}.

\bibitem{otherskklt}
K.~Choi, A.~Falkowski, H.~P. Nilles, and M.~Olechowski, ``{Soft supersymmetry
  breaking in KKLT flux compactification},''
  \href{http://dx.doi.org/10.1016/j.nuclphysb.2005.04.032}{{\em Nucl. Phys.}
  {\bfseries B718} (2005) 113--133},
\href{http://arxiv.org/abs/hep-th/0503216}{{\ttfamily arXiv:hep-th/0503216}}.

\bibitem{axiverse}
A.~Arvanitaki, S.~Dimopoulos, S.~Dubovsky, N.~Kaloper, and J.~March-Russell,
  ``{String Axiverse},''
  \href{http://dx.doi.org/10.1103/PhysRevD.81.123530}{{\em Phys. Rev.}
  {\bfseries D81} (2010) 123530},
\href{http://arxiv.org/abs/0905.4720}{{\ttfamily arXiv:0905.4720 [hep-th]}}.

\bibitem{bp}
R.~Bousso and J.~Polchinski, ``{Quantization of four form fluxes and dynamical
  neutralization of the cosmological constant},'' {\em JHEP} {\bfseries 0006}
  (2000) 006, \href{http://arxiv.org/abs/hep-th/0004134}{{\ttfamily
  arXiv:hep-th/0004134 [hep-th]}}.

\bibitem{kahlerstabilization}
T.~Banks and M.~Dine, ``{Coping with strongly coupled string theory},''
  \href{http://dx.doi.org/10.1103/PhysRevD.50.7454}{{\em Phys.Rev.} {\bfseries
  D50} (1994) 7454--7466},
  \href{http://arxiv.org/abs/hep-th/9406132}{{\ttfamily arXiv:hep-th/9406132
  [hep-th]}}.

\bibitem{retrofitted}
M.~Dine, J.~L. Feng, and E.~Silverstein, ``{Retrofitting O'Raifeartaigh models
  with dynamical scales},''
  \href{http://dx.doi.org/10.1103/PhysRevD.74.095012}{{\em Phys. Rev.}
  {\bfseries D74} (2006) 095012},
\href{http://arxiv.org/abs/hep-th/0608159}{{\ttfamily arXiv:hep-th/0608159}}.

\bibitem{dk}
M.~Dine and J.~Kehayias, ``{Discrete R Symmetries and Low Energy
  Supersymmetry},'' \href{http://dx.doi.org/10.1103/PhysRevD.82.055014}{{\em
  Phys. Rev.} {\bfseries D82} (2010) 055014},
\href{http://arxiv.org/abs/0909.1615}{{\ttfamily arXiv:0909.1615 [hep-ph]}}.

\bibitem{Breitenlohner1982}
P.~Breitenlohner and D.~Z. Freedman, ``{Positive Energy in anti-De Sitter
  Backgrounds and Gauged Extended Supergravity},''
\href{http://dx.doi.org/10.1016/0370-2693(82)90643-8}{{\em Phys. Lett.}
  {\bfseries B115} (1982) 197}.

\bibitem{Breitenlohner1982a}
P.~Breitenlohner and D.~Z. Freedman, ``{Stability in Gauged Extended
  Supergravity},''
\href{http://dx.doi.org/10.1016/0003-4916(82)90116-6}{{\em Ann. Phys.}
  {\bfseries 144} (1982) 249}.

\bibitem{haack}
M.~Berg, M.~Haack, and E.~Pajer, ``{Jumping Through Loops: On Soft Terms from
  Large Volume Compactifications},''
  \href{http://dx.doi.org/10.1088/1126-6708/2007/09/031}{{\em JHEP} {\bfseries
  0709} (2007) 031}, \href{http://arxiv.org/abs/arXiv:0704.0737}{{\ttfamily
  arXiv:arXiv:0704.0737 [hep-th]}}.

\bibitem{quevedoconloncorrections}
M.~Cicoli, J.~P. Conlon, and F.~Quevedo, ``{Systematics of String Loop
  Corrections in Type IIB Calabi-Yau Flux Compactifications},''
  \href{http://dx.doi.org/10.1088/1126-6708/2008/01/052}{{\em JHEP} {\bfseries
  0801} (2008) 052}, \href{http://arxiv.org/abs/arXiv:0708.1873}{{\ttfamily
  arXiv:arXiv:0708.1873 [hep-th]}}.

\bibitem{kehayiaswutoappear}
J.~Kehayias and W.~Wu. {To appear.}

\bibitem{conlon}
R.~Blumenhagen, J.~P. Conlon, S.~Krippendorf, S.~Moster, and F.~Quevedo,
  ``{SUSY Breaking in Local String/F-Theory Models},''
  \href{http://dx.doi.org/10.1088/1126-6708/2009/09/007}{{\em JHEP} {\bfseries
  09} (2009) 007},
\href{http://arxiv.org/abs/0906.3297}{{\ttfamily arXiv:0906.3297 [hep-th]}}.

\bibitem{conlonstability}
J.~P. Conlon and E.~Palti, ``{On Gauge Threshold Corrections for Local
  IIB/F-theory GUTs},''
  \href{http://dx.doi.org/10.1103/PhysRevD.80.106004}{{\em Phys. Rev.}
  {\bfseries D80} (2009) 106004},
\href{http://arxiv.org/abs/0907.1362}{{\ttfamily arXiv:0907.1362 [hep-th]}}.

\bibitem{bdgraesser}
T.~Banks, M.~Dine, and M.~Graesser, ``{Supersymmetry, axions and cosmology},''
  \href{http://dx.doi.org/10.1103/PhysRevD.68.075011}{{\em Phys. Rev.}
  {\bfseries D68} (2003) 075011},
\href{http://arxiv.org/abs/hep-ph/0210256}{{\ttfamily arXiv:hep-ph/0210256}}.

\bibitem{susskindovershoot}
B.~Freivogel, M.~Kleban, M.~Rodriguez~Martinez, and L.~Susskind,
  ``{Observational consequences of a landscape},'' {\em JHEP} {\bfseries 03}
  (2006) 039,
\href{http://arxiv.org/abs/hep-th/0505232}{{\ttfamily arXiv:hep-th/0505232}}.

\bibitem{wutoappear}
W.~Wu. {To appear.}

\end{thebibliography}\endgroup

\end{document}